\documentclass[aps,prd,reprint,superscriptaddress,twocolumn,nofootinbib]{revtex4-2}

\usepackage{amsmath,amssymb,amsthm,amstext}
\usepackage{graphicx}
\usepackage{color}
\usepackage{array, enumerate}
\usepackage{bm}
\usepackage{multirow}
\usepackage[breaklinks,colorlinks,citecolor=blue]{hyperref}
\usepackage{braket}
\usepackage{txfonts}
\usepackage{textcomp,gensymb}
\usepackage{enumitem}
\usepackage[dvipsnames]{xcolor}
\usepackage{orcidlink}
\usepackage[normalem]{ulem}
\usepackage[breaklinks,colorlinks,citecolor=blue]{hyperref}

\def\be{\begin{equation}}
\def\ee{\end{equation}}

\begin{document}
\title{Decomposition of Schwarzschild Green's Function}
\author{Junquan Su \orcidlink{0009-0008-1901-533X}}
\affiliation{Department of Astronomy, Tsinghua University, Beijing 100084, China}
\author{Neev Khera \orcidlink{0000-0003-3515-2859}}
\affiliation{Department of Astronomy, Tsinghua University, Beijing 100084, China}
\author{Marc Casals}
\affiliation{Institut f\"ur Theoretische Physik, Universit\"at Leipzig,\\ Br\"uderstra{\ss}e 16, 04103 Leipzig, Germany}
\affiliation{School of Mathematics and Statistics, University College Dublin, Belfield, Dublin 4, D04 V1W8, Ireland}
\affiliation{Centro Brasileiro de Pesquisas F\'isicas (CBPF), Rio de Janeiro, CEP 22290-180, Brazil}
\author{Sizheng Ma \orcidlink{0000-0002-4645-453X}}
\affiliation{Perimeter Institute for Theoretical Physics, Waterloo, ON N2L2Y5, Canada}
\author{Abhishek Chowdhuri \orcidlink{0000-0003-4474-790X}}
\author{Huan Yang \orcidlink{0000-0002-9965-3030}}
\email{hyangdoa@tsinghua.edu.cn}
\affiliation{Department of Astronomy, Tsinghua University, Beijing 100084, China}

\begin{abstract}
We present a formulation of the spherically decomposed Green's function for a Schwarzschild black hole, based on a decomposition into two components, $G^+$ and $G^-$, based on their large-frequency behaviour. While similar decompositions have been considered previously, here we systematically apply it to Schwarzschild spacetime and analyze its implications for the analytic structure of the Green's function in the complex-frequency plane.  
We show that both $G^+$ and $G^-$ possess branch cuts along the imaginary axis, which give rise to the direct part and the late-time tail, while the poles of $G^+$ correspond to the quasinormal mode spectrum. This allows us to identify a \textit{branch-cut direct part}, a quasinormal-mode contribution, and a late-time tail through contours adapted to different causal spacetime regions. This is in sharp contrast to Leaver's original formulation, where the prompt response is tied to a technically difficult large-arc contribution. We validate our decomposition with independent time-domain Regge-Wheeler simulations finding excellent agreement. Our results provide a practical and physically transparent framework for disentangling the distinct pieces of the Schwarzschild response, and offer a natural starting point for extensions to Kerr perturbations and non-linear ringdown physics.

\end{abstract}

\maketitle

\section{Introduction}

The Green's function of a black hole spacetime describes the signal sourced by a burst localized at a given point within the black hole spacetime. Depending on the nature of the equation, the corresponding signal may be a scalar wave, an electromagnetic wave, or a gravitational wave. If the underlying spacetime is Schwarzschild or Kerr, the Green's function can be obtained by, e.g.: numerically solving the wave equation on curved spacetime, either in 4-D \cite{Zenginoglu:2012xe} or after $\ell$-mode decomposition in $2$-D \cite{PhysRevD.89.084021,aruquipa2026greenfunctionsreggewheelerteukolsky};  by using quasinormal modes (QNMs), branch cut (BC) and a quasi-local analytical expansion \cite{Casals:2013mpa}; with WKB methods \cite{Dolan:2011fh,Yang:2012he,Yang:2013shb}. These results, together with fully mathematical approaches~\cite{Casals:2016qyj,harte2012caustics}, have shown that the Green's function exhibits a four-fold singular structure. In the context of gravitational wave lensing, such four-fold singular structure essentially defines  various types of strongly-lensed waveforms \cite{Dai:2017huk}.
This four-fold singular structure is a property of the full four-dimensional Green’s function.
Throughout this work, however, we use the term \emph{Green’s function} to refer to a single spherical 
$\ell$-mode of the four-dimensional Green’s function, which itself serves as the Green’s function of the corresponding two-dimensional reduced wave equation.

Mathematically, the time-domain Green's function can be computed by performing a Fourier transform of the frequency domain Green's function, as illustrated by Leaver in \cite{Leaver1986} for the Schwarzschild background, which in turn gives three individual components after applying a complex contour technique: the direct part, the QNMs, and the tail. Physically, the \emph{direct part} should correspond  to the part of signal directly emitted towards an observer, although generally it should contain a piece on the light cone and another piece within the light cone. QNMs, being the poles of the Green's function, are effectively the waves winding around the black hole and eventually escaping to the observer. The tail, on the other hand, is generated by scattering waves off the $1/r$ gravitational potential at large distances. Therefore it seems both mathematically plausible and physically motivated to decompose the time-domain Green's function into these three parts, which should facilitate analytical understanding of gravitational wave generation in various compact binary systems relevant for gravitational wave detections \cite{Berti:2025hly}.

However, despite being physically intuitive, there are still serious issues associated with this decomposition of the Green's function. First, the direct part defined on the ``large arc" of the contour integration is difficult to compute in a convergent limit. Second, both the summation of QNMs with different overtones and the tail term generically diverge before a certain starting time \cite{PhysRevD.86.024021}. Since the total causal Green's function remains finite at all time, this means that the divergences of QNMs and the tail may cancel each other or have to be canceled by a counter-diverging term in the direct part. This theoretical problem poses a serious challenge on the program of black hole spectroscopy, which aims to measure individual linear QNMs in the ringdown stage of a binary black hole merger.
Strictly speaking, before the cut-off time that defines the convergent zone of QNM sum, searching for modes with various overtones is theoretically ill defined.

The divergent sum problem of QNMs essentially comes because it should not be applied to the entire domain of spacetime. Physically, a burst source placed outside the light ring of a black hole should only start exciting the QNMs once the ingoing waves hit the light ring. Therefore, it is reasonable to impose a cut-off time on the QNM signal, as illustrated in a toy problem in $1+1$ dimensions \cite{Chavda:2024awq}. This proposal has been made rigorous for Schwarzschild spacetime in \cite{DeAmicis:2025xuh}, where the Green's function is restructured so that the QNM contribution has a natural cut-off time within the contour integration approach, giving rise to ``dynamically exited modes" in the ringdown stage. 

Moreover, it has been shown that for Schwarzschild--de Sitter black holes,
both the quasinormal-mode and direct contributions can be obtained in a
self-consistent manner~\cite{arnaudo2025quasinormalmodescompletemode}.
In Schwarzschild--de Sitter spacetime, the presence of a cosmological
horizon leads to a discrete set of imaginary-axis frequencies, commonly
referred to as \emph{Matsubara modes}.
In Refs.~\cite{arnaudo2025quasinormalmodescompletemode,arnaudo2025priceslawquasinormalmodes},
the Green’s function is decomposed into two components, $G^{+}$ and
$G^{-}$, with different analytic structures. It is shown that the direct part (which is the Schwarzchild-de Sitter equivalent of the \emph{branch-cut direct part} here in Schwarzschild) arises from the residues of the
Matsubara modes of $G^{+}$, together with the zero mode and the
lower-half-plane Matsubara modes of $G^{-}$.
The late-time tail is likewise generated by contributions from the
Matsubara spectrum.

Refs.~\cite{arnaudo2025priceslawquasinormalmodes} points out that, in the limit of a
vanishing cosmological constant $\Lambda \to 0$, the discrete Matsubara modes
coalesce into a continuous structure, giving rise to the BC that
accounts for the late-time tail in Schwarzschild spacetime.
Motivated by Refs.~\cite{arnaudo2025quasinormalmodescompletemode,arnaudo2025priceslawquasinormalmodes}, it is natural to expect that as $\Lambda \to 0$,
the Matsubara modes of $G^{+}$ and $G^{-}$ on both the positive imaginary axis
(PIA) and the negative imaginary axis (NIA) merge into BCs along the
corresponding axes. 
In this work, we therefore develop a framework to separately identify the
direct part, the quasinormal-mode contribution, and the tail of the
gravitational-perturbation Green’s function in Schwarzschild spacetime.
As in Refs.~\cite{arnaudo2025priceslawquasinormalmodes}, we decompose the Schwarzschild
Green’s function into $G^{+}$ and $G^{-}$. Consistent with our expectation
outlined above, we find that, instead of a discrete set of Matsubara modes on
the imaginary axis, BCs appear along the imaginary-frequency axis in
Schwarzschild spacetime, and that these BCs contribute to both the
direct part and the tail.

To validate this decomposition, we numerically compute waveforms sourced
by a narrow Gaussian profile, serving as an approximation to a
delta-function source, using a time-domain Regge--Wheeler solver.
We demonstrate that the numerical waveforms agree with the individual
contributions reconstructed from the Green’s function at different
observation times.
Physically, our results are consistent with the picture 
in Ref.~\cite{arnaudo2025quasinormalmodescompletemode,arnaudo2025priceslawquasinormalmodes} in the
small–cosmological-constant limit, while highlighting that, in
Schwarzschild spacetime, the tail contribution remains an intrinsic and
essential feature.

We emphasize that the \emph{branch-cut direct part} identified in this work is
conceptually different from the direct part associated with the large-arc
contribution in Leaver’s original contour-integral formulation
\cite{Leaver1986}. Although previous studies have successfully modeled the
quasinormal-mode and tail contributions in Schwarzschild spacetime
\cite{Casals:2013mpa}, the direct part arising from the large-arc
contribution in Leaver’s contour is notoriously difficult to evaluate in
practice.
By contrast, the branch-cut direct part obtained in our decomposition is
computationally more tractable. For this reason, we argue that a spectral
decomposition based on the branch-cut direct part provides a more natural and
practical framework for decomposing the Green’s function in Schwarzschild
spacetime.



The structure of this paper is as follows. In Sec.~\ref{sec:theo}, we present the theoretical framework for the decomposition of the Green’s function in Schwarzschild spacetime and describe the construction of integration contours in different regimes. Section~\ref{Implementation} is devoted to the numerical implementation of our method. In particular, Sec.~\ref{subsec:MST} introduces the Mano–Suzuki–Takasugi (MST) formalism employed in our calculations, while Sec.~\ref{subsectionDP} explains the computation of the direct part of the Green’s function. In Sec.~\ref{tailQNM}, we demonstrate how the tail and quasinormal-mode (QNM) waveforms are computed. Section~\ref{Simulation} briefly describes the numerical procedure used to simulate the Green’s function. Finally, the results are presented and discussed in Sec.~\ref{SEC:Result}.

In this paper we work in units $c=G=M=1$. Quantities with an overbar indicate dimensionless variables rescaled by appropriate powers of the event-horizon radius $r_h=2M$.



\section{Theoretical framework}\label{sec:theo}

Linear perturbations of the Schwarzschild spacetime can be reduced to a one-dimensional wave equation. 
In this work, we focus on the odd-parity (axial) sector, which is governed by the Regge–Wheeler equation~\cite{PhysRev.108.1063}. 
The even-parity (polar) perturbations satisfy the Zerilli equation; however, the corresponding homogeneous solutions can be obtained from the Regge–Wheeler solutions via the supersymmetric (SUSY) relation~\cite{chandrasekhar1975equations, chandrasekhar1998mathematical}. 
Therefore, restricting our analysis to the odd-parity sector does not entail any essential loss of generality at the level of homogeneous solutions. 
The master variable $\psi_{\ell}$ satisfies
\begin{equation}
\left[
-\frac{\partial^2}{\partial t^2}
+ \frac{\partial^2}{\partial r_*^2}
- V_{\rm RW}(r)
\right]\psi_{\ell }(t,r)
= 0 ,
\end{equation}
where $r_*$ is the radial `tortoise coordinate' defined as $r_* = r+2M\ln \left( \frac{r}{2M} -1\right)$ and $V_{\rm RW}(r)$ is the Regge–Wheeler potential 
\begin{equation}
V_{\rm RW}(r)
=
\left(1-\frac{2M}{r}\right)
\left[
\frac{\ell(\ell+1)}{r^2}
-\frac{6M}{r^3}
\right].
\end{equation}
In the frequency domain, the Regge–Wheeler equation takes the form
\begin{equation}
\left[
\frac{d^2}{d r_*^2}
+\omega^2
- V_{\rm RW}(r)
\right]\psi_{\ell }(\omega,r)
= 0 .
\label{FreqDomainRW}
\end{equation}

The ingoing (IN) and upgoing (UP) solutions of the homogeneous
Regge–Wheeler equation are defined by their asymptotic behaviors,
\begin{align}
R^{\mathrm{in}}_{\omega}(r)
&\sim
\begin{cases}
 e^{-i \omega r_*},
& r_* \to -\infty, \\[4pt]
A^{\mathrm{ref}}_{\mathrm{in}}(\omega)\, e^{ i \omega r_*}
+
A^{\mathrm{inc}}_{\mathrm{in}}(\omega)\, e^{-i \omega r_*},
& r_* \to +\infty ,
\end{cases}
\\[6pt]
R^{\mathrm{up}}_{\omega}(r)
&\sim
\begin{cases}
A^{\mathrm{inc}}_{\mathrm{up}}(\omega)\, e^{ i \omega r_*}
+
A^{\mathrm{ref}}_{\mathrm{up}}(\omega)\, e^{-i \omega r_*},
& r_* \to -\infty , \\[4pt]
e^{ i \omega r_*},
& r_* \to +\infty .
\end{cases}
\end{align}

The downgoing (DOWN) solution is defined as the homogeneous solution
satisfying $ R^{\mathrm{down}}_{\omega}(r) \sim e^{-i\omega r_*}, $ as $r_* \to +\infty 
$.
Here, all radial solutions and amplitudes are evaluated at fixed angular index $\ell$, and the $\ell$ dependence is suppressed for brevity.

{The Fourier-domain Green's function associated with the Regge--Wheeler equation can be written as
\begin{equation}
\tilde{G}(\omega; r, r')
=
\frac{
R^{\mathrm{up}}_\omega(r)\,
R^{\mathrm{in}}_\omega(r')
}
{W(\omega)} \,\Theta(r - r')
+
\frac{
R^{\mathrm{up}}_\omega(r')\,
R^{\mathrm{in}}_\omega(r)
}
{W(\omega)} \,\Theta(r' - r)
\,,
\end{equation}
where $\Theta$ denotes the Heaviside step function. In most situations of interest, \( r > r' \), and therefore only the first term contributes. Furthermore, since observers are typically located far from the black hole, we focus on the case where \( r_* > 0 \) in this paper.
The quantity
\begin{equation}
W(\omega)
\equiv
R^{\mathrm{up}}_\omega \frac{d R^{\mathrm{in}}_\omega}{d r_*}
-
R^{\mathrm{in}}_\omega \frac{d R^{\mathrm{up}}_\omega}{d r_*}
= 2 i \omega A^{\mathrm{inc}}_{\mathrm{in}}
\end{equation}
is the Wronskian of the IN and UP solutions, which is independent of the radial coordinate.
}

Using the standard scattering relation
\begin{equation}
R^\text{in}_\omega(r)
=
A^\text{inc}_\text{in}(\omega)\, R^\text{down}_\omega(r)
+
A^\text{ref}_\text{in}(\omega)\, R^\text{up}_\omega(r),
\end{equation}
the Green's function can be naturally decomposed into two contributions,
\begin{equation}
\tilde{G}(\omega; r,r^\prime)
=
\tilde{G}^{+}(\omega; r,r^\prime)
+
\tilde{G}^{-}(\omega; r,r^\prime).
\label{eqn:GF_Decomposition}
\end{equation}

These two components are explicitly given by
\begin{align}
\tilde{G}^{+}(\omega; r,r^\prime)
&=
\frac{
A^\text{ref}_\text{in}(\omega)\,
R^\text{up}_\omega(r)\,
R^\text{up}_\omega(r^\prime)
}
{2 i \omega\, A^\text{inc}_\text{in}(\omega)},
\label{Gplus}
\\[1ex]
\tilde{G}^{-}(\omega; r,r^\prime)
&=
\frac{
R^\text{up}_\omega(r)\,
R^\text{down}_\omega(r^\prime)  
}
{2 i \omega}.
\label{Gminus}
\end{align}
Hereafter, we denote $\tilde{G}^{+}(\omega; r,r^\prime)$ and
$\tilde{G}^{-}(\omega; r,r^\prime)$ simply by $G^{+}$ and $G^{-}$.

The QNMs of the Green’s function correspond to the
zeros of the incident amplitude $A^\mathrm{inc}_\mathrm{in}(\omega)$,
which appears in the Wronskian.
Since $A^\mathrm{inc}_\mathrm{in}(\omega)$ enters only in
$G^{+}$, it follows that only $G^{+}$ contributes to the
QNM spectrum, while $G^{-}$ is free of QNM poles.
Regarding the BC structure, the IN solution does not contribute
to any BC, as evidenced by its Jaff\'e series representation [Eq.~\ref{Jaffe}],
which is analytic along the imaginary axis.
In contrast, the UP solution expressed through the Leaver $U$-series [Eq.~\ref{LeaverU}]
contains the confluent hypergeometric function $U$, which possesses a
BC along the NIA.
By virtue of the relation 
$R^\mathrm{down}_{\omega} = [R^\mathrm{up}_{\omega^*}]^{*}$,
the DOWN solution exhibits a corresponding BC along the PIA. 

For $G^{+}$, the UP solution contributes a BC on the NIA
through both the radial function and the Wronskian, while the reflection
amplitude $A^\mathrm{ref}_\mathrm{in}(\omega)$ introduces an additional
BC on the PIA \footnote{
The Wronskian of the IN and DOWN solutions,
$W[R^\mathrm{in}_\omega, R^\mathrm{down}_\omega]$, which is likewise
independent of the radial coordinate $r$, is given by
$W = -2 i \omega A^\mathrm{ref}_\mathrm{in}(\omega)$.
Since the DOWN solution possesses a BC on the PIA, the Wronskian
inherits this BC, implying that
$A^\mathrm{ref}_\mathrm{in}(\omega)$ also has a BC along the PIA.}.
For $G^{-}$, the UP solution contributes to the BC on the
NIA, whereas the DOWN solution contributes to the BC on the PIA. 

In summary, both $G^{+}$ and $G^{-}$ individually possess
BCs on both the PIA and NIA. When these two components are
combined to form the full Green’s function, the contributions from the
PIA cancel exactly, leaving a net BC only along the NIA.


\begin{figure*}[t]
  \centering
  \includegraphics[width=1.0\textwidth]{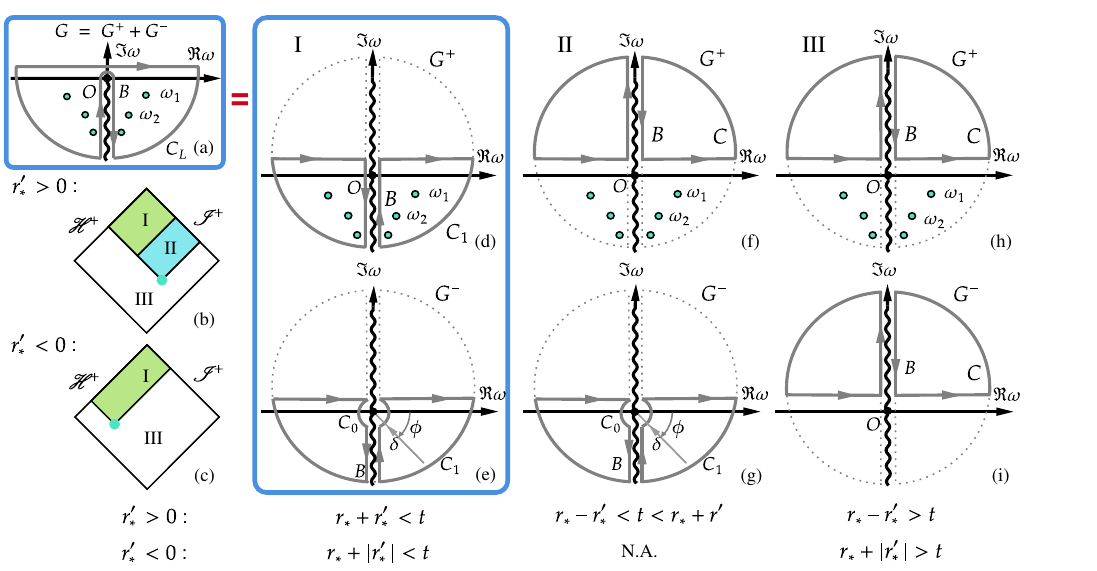}
\caption{The decomposition of the Schwarzschild black hole Green's function into the regions I, II and III, as depicted in the penrose diagrams on the left. In region I, both $G^+$ and $G^-$ contours are closed on bottom half plane, making the result identical to the traditional contour integration of the total Green's function depicted in the top left. In region II, the $G^+$ contour is instead closed on the top half while the $G_-$ contour is closed in the bottom half. We note that if $r_*^\prime<0$, there is no region II. In region III both $G+$ and $G^-$ contours are closed in the top half, and the total greens function will vanish. Here, we use \( C_L \) to denote the large arc in Leaver's traditional contour, and \( C \), \( C_1 \), and \( C_0 \) to denote the large arcs in the upper half-plane, lower half-plane, and the small circular detour, respectively. \( B \) represents the branch cuts.}
  
  
  \label{fig:Contour}
\end{figure*}

The time-domain Green’s function is obtained by a Fourier transform of its frequency-domain counterpart,
\begin{equation}\label{eq:Fourier G}
G(t; r,r^\prime) = \frac{1}{2\pi} \int_{-\infty + ic}^{+\infty + ic} d\omega \;
\tilde{G}(\omega; r,r^\prime)\, e^{-i\omega t}.
\end{equation}
where $c>0$.
This integral can be evaluated by deforming the integration contour in the complex $\omega$ plane, as illustrated in Fig.~\ref{fig:Contour}.

Following the spacetime decomposition introduced in Refs.~\cite{arnaudo2025quasinormalmodescompletemode,kuntz2025greenfunctionposchltellerpotential}, we partition the spacetime using different null rays.
When the source is located at $r^\prime_*>0$, the spacetime is separated by an ingoing null ray, an outgoing null ray, and a null ray reflected by the Regge--Wheeler potential, as illustrated in the Penrose diagram in the (b) panel of Fig.~\ref{fig:Contour}. The spacetime is thereby divided into three regions, labeled I, II, and III. In addition, when the source is located at $r^\prime_*<0$, the spacetime is instead divided into two regions, I and III, by an ingoing and an outgoing null ray, since a reflected null ray cannot reach an observer on the right side of the Regge--Wheeler potential, as illustrated in the Penrose diagram in the (c) panel of Fig.~\ref{fig:Contour}.
We stress that there is no ``direct part" of the Green's function in this case, and the prompt signal on the light cone is solely contributed by the summation of quasinormal modes. The latter point have been illustrated by using WKB methods in Schwarzschild \cite{Dolan:2011fh} and in Kerr \cite{Yang:2013shb}.

When $t > |r_*| + |r_*^\prime|$ (Region I), we employ the standard Leaver contour shown as panel (a) in Fig.~\ref{fig:Contour}.
In this regime, different segments of the contour give rise to distinct physical contributions.
The BC of $G^+ + G^-$ 
along the NIA generates the late-time power-law tail, while the poles of the Green’s function correspond to the QNM contribution~\cite{Leaver1986}.

For the intermediate time range $r_* - r_*^\prime < t < r_* + r_*^\prime,  \, (r_*^\prime>0)$ (Region II), corresponding to the direct part of the signal in the case $r'_*>0$, it is convenient to decompose the Green’s function as $G = G^+ + G^-$ and evaluate the two components separately.
In the large-\(|\omega|\) limit, these components behave as \( G^{+} \sim e^{i\omega (r_* + r_*^\prime)} \) and \( G^{-} \sim e^{i\omega (r_* - r_*^\prime)} \), since the Regge-Wheeler potential term in [Eq.~\ref{FreqDomainRW}] can be ignored for sufficiently large \(|\omega|\).
The Fourier factor $e^{-i\omega t}$ then suppresses the contribution from the large arc in the appropriate half-plane.

For the \( G^+ \) component, we construct a contour in the upper half of the complex \( \omega \) plane, as shown in panel (f) of Fig.~\ref{fig:Contour}. Since \( G^+ \) possesses BCs on both the PIA and NIA, a simple semicircular contour would necessarily cross a BC. To avoid this, we introduce two quarter-circle arcs on the left- and right-hand sides of the imaginary axis, with their radial edges taken to lie infinitesimally close and parallel to the imaginary axis. In this construction, no poles are enclosed, and only the BC on the PIA contributes to the waveform; in particular, no QNM contribution is present.



The \( G^- \) contribution is evaluated using a similar contour in the lower half-plane. In this case, the discontinuity of \( G^- \) exhibits a divergence as \( \omega \to 0 \) along the imaginary axis, which arises from the properties of the confluent hypergeometric function in [Eq.~\ref{LeaverU}] as well as the factor of \( \omega \) in the numerator. To avoid this issue, a small circular detour is introduced to isolate and regulate the singular behavior, as illustrated in panel (g) of Fig.~\ref{fig:Contour}.   By contrast, in the traditional contour formulation, the full Green's function \( G \) vanishes as \( \omega \to 0 \) along the imaginary axis (e.g.,~\cite{PhysRevD.92.124055}), and no such detour is required.

We note that the contours in region II are not applicable for \( r^\prime_* < 0 \). This can be understood from the Penrose diagram shown in panel (c) of Fig.~\ref{fig:Contour}, where region II is absent when \( r^\prime_* < 0 \). For \( t < |r_*| + |r^\prime_*| = r_* - r^\prime_* \), which can be interpreted as a causal boundary, the contour for region III described in the following paragraph should be used. Consequently, there is no contribution from the direct part to the Green’s function waveform when \( r^\prime_* < 0 \).

When $t < r_* - r_*^\prime$ (Region III), the contours for both $G^+$ and $G^-$ are taken in the upper half-plane, as shown in contour~III of Fig.~\ref{fig:Contour}.
The contribution from the large arc is again exponentially suppressed by the Fourier factor.
Moreover, the BC contributions from $G^+$ and $G^-$ cancel exactly when the two components are summed, yielding a vanishing Green’s function.
From a physical perspective, this is a direct consequence of causality.

In summary, for late times, we use the standard contour for $G^+ + G^-$ and recover the usual QNM and late-time tail contributions.
However, for the early-time signal, the traditional contour representation is of limited practical use, since a direct evaluation of the large-arc contribution is technically difficult.
For this reason, we instead employ the above decomposition of the Green’s function and construct separate contours for $G^+$ and $G^-$.
This strategy is inspired by the approach introduced in Ref.~\cite{arnaudo2025quasinormalmodescompletemode}, suitably adapted to the present context.




\section{Implementation} \label{Implementation}

\subsection{MST Method}
\label{subsec:MST}
To evaluate the contour integrals appearing in the frequency-domain
Green’s function, it is necessary to compute the Regge–Wheeler radial
solutions and their asymptotic amplitudes with high accuracy over the
complex frequency plane. For this purpose, we employ the
MST method, which provides accurate
representations of the Regge–Wheeler solutions and their asymptotic amplitudes for generic complex frequencies.

In this work, our focus is on the evaluation of
the frequency-domain Green’s function and its individual contributions,
including the direct part, QNMs and tail to the time-domain waveform. Although the \texttt{ReggeWheeler} module in the
\texttt{BlackHolePerturbationToolkit} provides a well-developed
MST-based implementation in \texttt{Mathematica}, we develop an
independent and dedicated numerical implementation of the MST formalism
in \texttt{Julia}, following the framework described in
Ref.~\cite{PhysRevD.92.124055}. This implementation enables systematic
and numerically stable evaluations of the Regge–Wheeler radial
solutions, their Wronskians, and the associated asymptotic amplitudes
required for constructing the Green’s function.

While the present work is restricted to the computation and analysis of
the Green’s function itself, the \texttt{Julia}-based implementation is designed
with future extensions in mind. In particular, it provides a flexible
foundation for applications requiring large-scale and
high-performance evaluations, such as time-domain waveform
reconstruction from decomposed Green’s functions. Within the same
framework, we have also implemented complementary radial
representations, including the Jaff\'e series and the Leaver
$U$-series, which are employed in different frequency regimes to
optimize numerical accuracy and efficiency.

Within the MST formalism, the UP solution of the Regge–Wheeler equation can be expressed as the following series
\begin{align}
   R^\text{up}_\omega (r)
  &= 
  \left( 1 - \frac{\bar{\omega}}{z}  \right)^{-i\bar{\omega}}
  N_s^{\text{up}} e^{iz} (-2iz)^{\nu+1}  
  \nonumber \\
  &\times
  \sum_{n=-\infty}^{\infty} \frac{\Gamma(n+\nu+1+s-i\bar{\omega})\Gamma(n+\nu+1-i\bar{\omega})}{\Gamma(n+\nu+1-s+i\bar{\omega})\Gamma(n+\nu+1+i\bar{\omega})} \nonumber \\
  &\times
  a_n^\nu (-2iz)^n U(n+\nu+1-i\bar{\omega}, 2n+2\nu+2, -2iz),
\end{align}
where $\bar{\omega} \equiv 2M\omega$, $z \equiv \omega r = \bar{\omega}\,\bar{r}$,
and $\bar{r} \equiv r/(2M)$.
The parameter $\nu$ denotes the renormalized angular momentum,
which is determined by solving the continued-fraction equation
given in [Eq.~\eqref{eq:nu_cf}].
In this work, we explicitly compute only the odd-parity
gravitational perturbations, corresponding to spin weight $s=2$.

Here, $U(a,b,z)$ denotes the irregular confluent hypergeometric function~\cite{DLMF}.
In our numerical implementation, the function $U$ and the function $\Gamma$ are evaluated using arbitrary-precision complex arithmetic provided by the \texttt{Nemo.jl} package~\cite{Nemo}.

The normalization factor $N_s^{\mathrm{up}}$ is given by
\begin{equation}
N_s^{\text{up}} = \frac{1}{2} e^{-\pi \bar{\omega}} \frac{\Gamma(\nu+1-s+i\bar{\omega})\Gamma(\nu+1+i\bar{\omega})}{\Gamma(\nu+1+s-i\bar{\omega})\Gamma(\nu+1-i\bar{\omega})} e^{-i \frac{\pi}{2} (\nu+1)}.
\end{equation}



The expansion coefficients $a_n^\nu$ are determined by the three-term recurrence relation
\begin{equation}
    \alpha^\nu_n a_{n+1} + \beta^\nu_n a_n + \gamma^\nu_n a_{n-1} = 0, \qquad n \in \mathbb{Z},
\end{equation}
where $\alpha_n^\nu$, $\beta_n^\nu$, and $\gamma_n^\nu$ are given by
\begin{align}
\alpha_n^\nu = &\frac{i\bar{\omega}(n+\nu+1-i\bar{\omega})(n+\nu+1+s-i\bar{\omega})}{(n+\nu + 1)(2n+2\nu+3)(n+\nu+1+s+i\bar{\omega})^{-1}}
\label{MSTalpha}
\\
\beta_n^\nu =& \Lambda - (n+\nu)(n+\nu+1) - 2\bar{\omega}^2 -  \frac{\bar{\omega}^2 ( \bar{\omega}^2 + s^2 )}{(n+\nu + 1)(n+\nu)}
\label{MSTbeta}
\\
\gamma_n^\nu =& -\frac{i\bar{\omega}(n+\nu+i\bar{\omega})(n+\nu-s+i\bar{\omega})(n+\nu-s-i\bar{\omega})}{(2n+2\nu-1)(n+\nu)}
\label{MSTgamma}
\end{align}
with $\Lambda = \ell(\ell+1)$. The overall normalization of the series is fixed by choosing $a_0^\nu = 1$. The sequence $\{a_n^\nu\}$ can be constructed via the
continued-fraction method recursively once the renormalized angular momentum $\nu$ has been determined by [Eq.~\eqref{eq:nu_cf}].

In the MST framework, the asymptotic amplitudes of the IN and UP solutions can be computed in terms of the MST coefficients.
In particular, the incident, reflected, and transmitted amplitudes of the IN solution,
$A^{\mathrm{inc}}_{\mathrm{in}}(\omega)$,
$A^{\mathrm{ref}}_{\mathrm{in}}(\omega)$,
$A^{\mathrm{tra}}_{\mathrm{in}}(\omega)$,
as well as the transmitted amplitude of the UP solution,
$A^{\mathrm{tra}}_{\mathrm{up}}(\omega)$,
are given by
\begin{equation}
A^\text{inc}_\text{in} (\omega)
= \left( K_{\nu} - i e^{-i\pi \nu} \frac{\sin\left(\pi(\nu+i\bar{\omega})\right)}{\sin\left(\pi(\nu-i\bar{\omega})\right)} K_{-\nu-1} \right) \check{A}_{+}^{\nu} e^{-i\bar{\omega} \ln \bar{\omega}}
\end{equation}
\begin{equation}
A^\text{ref}_\text{in} (\omega)
= \left( K_{\nu} + i e^{i\pi \nu} K_{-\nu-1} \right) (2i)^s A_{-}^{\nu} e^{i\bar{\omega} \ln \bar{\omega}}
\end{equation}
\begin{align}
A^\text{tra}_\text{in} (\omega)
&= 
\frac{\Gamma(1-s-2i\bar{\omega})}{\Gamma(-\nu-i\bar{\omega})\Gamma(1+\nu-i\bar{\omega})}
e^{i\bar{\omega}} 
\nonumber \\
&\times
\sum_{n=-\infty}^{\infty} 
a_n^\nu 
\frac{
         \Gamma(-n-\nu+s-i\bar{\omega})
     }
     {  
         \Gamma(1-2i\bar{\omega})
         \left[
                 \Gamma(n+\nu+s+1-i\bar{\omega})
         \right]^{-1}
     },
\end{align}
\begin{equation}
A^\text{tra}_\text{up} (\omega)
= (2i)^s A_{-}^{\nu} e^{i\bar{\omega} \ln \bar{\omega}},
\end{equation}
where
\begin{align}
A_{-}^{\nu} 
\equiv &
(2i)^{-s} e^{-\frac{\pi}{2}\bar{\omega}} e^{-i\frac{\pi}{2}(\nu+1)} 2^{-1+i\bar{\omega}} \\
\nonumber 
&\times 
\sum_{n=-\infty}^{\infty} \frac{(\nu+1+s-i\bar{\omega})_n (\nu+1-i\bar{\omega})_n}{(\nu+1-s+i\bar{\omega})_n (\nu+1+i\bar{\omega})_n} a_n^\nu,
\end{align}

\begin{align}
\check{A}_{+}^{\nu} 
\equiv &
e^{-\pi \bar{\omega} / 2} e^{i \frac{\pi}{2} (\nu + 1)} 2^{-1 - i \bar{\omega}} \frac{\Gamma(\nu + 1 - s + i \bar{\omega}) \Gamma(\nu + i \bar{\omega} + 1)}{\Gamma(\nu + 1 + s - i \bar{\omega}) \Gamma(\nu - i \bar{\omega} + 1)} 
\nonumber \\
&\times
\sum_{n=-\infty}^{\infty} (-1)^n \frac{\Gamma(n + \nu - i \bar{\omega} + 1 + s)}{\Gamma(n + \nu + i \bar{\omega} + 1 - s)} a_n^\nu,
\end{align}
and
\begin{widetext}
\begin{align}
K_{\nu} &
= 
\frac{
       e^{i\bar{\omega}}(2\bar{\omega})^{s-\nu}2^{-s}
       \Gamma(1-s-2i\bar{\omega})
       \Gamma(2\nu+2)
    }
{
      \Gamma(\nu+1-s+i\bar{\omega})
      \Gamma(\nu+1+i\bar{\omega})
      \Gamma(\nu+1+s+i\bar{\omega})
} 
\left( 
      \sum_{n=0}^{\infty} 
      \frac{
             \Gamma(n+2\nu+1)
           }
           {
              n!
           }
      \frac{
             \Gamma(n+\nu+1+s+i\bar{\omega})
             \Gamma(\nu+1+i\bar{\omega})
           }
           {
             \Gamma(n+\nu+1-s-i\bar{\omega})
             \Gamma(\nu+1-i\bar{\omega})} 
             a_n^\nu
\right) \nonumber \\
&\times 
\left( 
         \sum_{n=-\infty}^{0} 
         \frac{1}{(-n)!(2\nu+2)_n} 
         \frac{
                (\nu+1+s-i\bar{\omega})_n
                (\nu+1-i\bar{\omega})_n
              }
              {
                (\nu+1-s+i\bar{\omega})_n
                (\nu+1+i\bar{\omega})_n
              } 
         a_n^\nu
\right)^{-1}
\end{align}
Here $(a)_n \equiv \Gamma(a+n)/\Gamma(a)$ denotes the Pochhammer symbol.
\end{widetext}


In our MST implementation, the renormalized angular momentum $\nu$ is determined by using a dedicated numerical continuation strategy.
The quantity $\nu$ arises in the Mano--Suzuki--Takasugi formalism as a deformation of the angular momentum index $\ell$, and is defined implicitly by the requirement that the series coefficients $a_n^\nu$ of the MST expansion satisfy a three-term recurrence relation,
\begin{equation}
\alpha_n^\nu a_{n+1}^\nu + \beta_n^\nu a_n^\nu + \gamma_n^\nu a_{n-1}^\nu = 0 ,
\label{eq:MST_recursion}
\end{equation}
which admits a minimal solution in both limits $n \to +\infty$ and $n \to -\infty$.
This condition leads to a characteristic equation for $\nu$ in the form of an infinite continued fraction
\begin{equation}
    \beta^\nu_n + \alpha^\nu_n R_{n+1} + \gamma^\nu_n L_{n-1} = 0,
    \label{eq:nu_cf}
\end{equation}
where $R_{n} \equiv  a^\nu_n /a^\nu_{n-1}$ and $L_{n} \equiv  a^\nu_n /a^\nu_{n+1}$. Using the recurrence relation, $R^\nu_n$ and $L^\nu_n$ can be expressed as the following continued fractions
\begin{align}
R_n = 
\cfrac{ -\gamma_n^\nu}{
\beta_n^\nu
-
\cfrac{\alpha_{n}^\nu \gamma_{n+1}^\nu}{
\beta_{n+1}^\nu
-
\cfrac{\alpha_{n+1}^\nu \gamma_{n+2}^\nu}{
\beta_{n+2}^\nu - \cdots
}}},
\quad
L_n = 
\cfrac{ -\alpha_n^\nu}{
\beta_n^\nu
-
\cfrac{\alpha_{n-1}^\nu \gamma_{n}^\nu}{
\beta_{n-1}^\nu
-
\cfrac{\alpha_{n-2}^\nu \gamma_{n-1}^\nu}{
\beta_{n-2}^\nu - \cdots
}}}.
\end{align}
[Eq.~\eqref{eq:nu_cf}] implicitly defines $\nu$ as a function of the complex frequency $\omega$.
The explicit expressions for the recurrence coefficients $\alpha_n^\nu$, $\beta_n^\nu$, and $\gamma_n^\nu$ follow [Eq.~\eqref{MSTalpha}], [Eq.~\eqref{MSTbeta}] and [Eq.~\eqref{MSTgamma}].

Numerically, [Eq.~\eqref{eq:nu_cf}] is solved along a continuous trajectory in the complex $\omega$ plane.
At each frequency step, the value of $\nu$ obtained at the previous frequency is used as the initial guess for the root-finding procedure.
This continuation strategy allows us to robustly track the physically relevant branch of $\nu$ for frequencies along the imaginary axis as well as along the small circular contour.
The same procedure is employed in our computations of the quasinormal-mode and BC tail contributions.


In this subsection, we have summarized the key MST formulae relevant to our analysis. While the MST method provides a highly accurate and robust framework, it is not always the most efficient choice for every component of the computation. In particular, for the radial IN solution we employ the Jaff\'e series representation [Eq.~\eqref{Jaffe}], while in the high-frequency regime we use the Leaver $U$-series
[Eq.~\eqref{LeaverU}] to evaluate the UP solution. These approaches do not require the computation of the renormalized angular momentum $\nu$, making them simpler to implement and computationally more efficient. Further details are discussed in Sec.~\ref{tailQNM}.


\subsection{Direct part} \label{subsectionDP}

To evaluate the direct contribution to the waveform via contour integration, we compute the UP and DOWN solutions of the Regge--Wheeler equation, together with their asymptotic amplitudes, along the  imaginary-frequency axis and the small circular contour. The MST method provides a suitable and robust framework for carrying out these calculations.

In the numerical implementation, we slightly shift the BC to the right of the imaginary axis by an amount $\epsilon = 10^{-80}$, so that the Green’s function is effectively evaluated on the right-hand side of the BC. Owing to the complex-conjugation symmetry of the Regge--Wheeler radial solutions,
\begin{equation}
    R^{\mathrm{in/up/down}}_{\omega}(r)
    =
    \left[
        R^{\mathrm{in/up/down}}_{-\omega^{*}}(r)
    \right]^{*},
    \thinspace
    W(\omega)
    =
    \left[
        W(-\omega^{*})
    \right]^{*}.
    \label{ConjSym}
\end{equation}
the corresponding quantities on the left-hand side of the BC can be
obtained by complex conjugation of those evaluated on the right-hand
side. Similar symmetry relations also hold for the asymptotic amplitudes
and for the frequency-domain Green’s functions.

\begin{figure}[t]
  \centering
  \includegraphics[width=\linewidth]{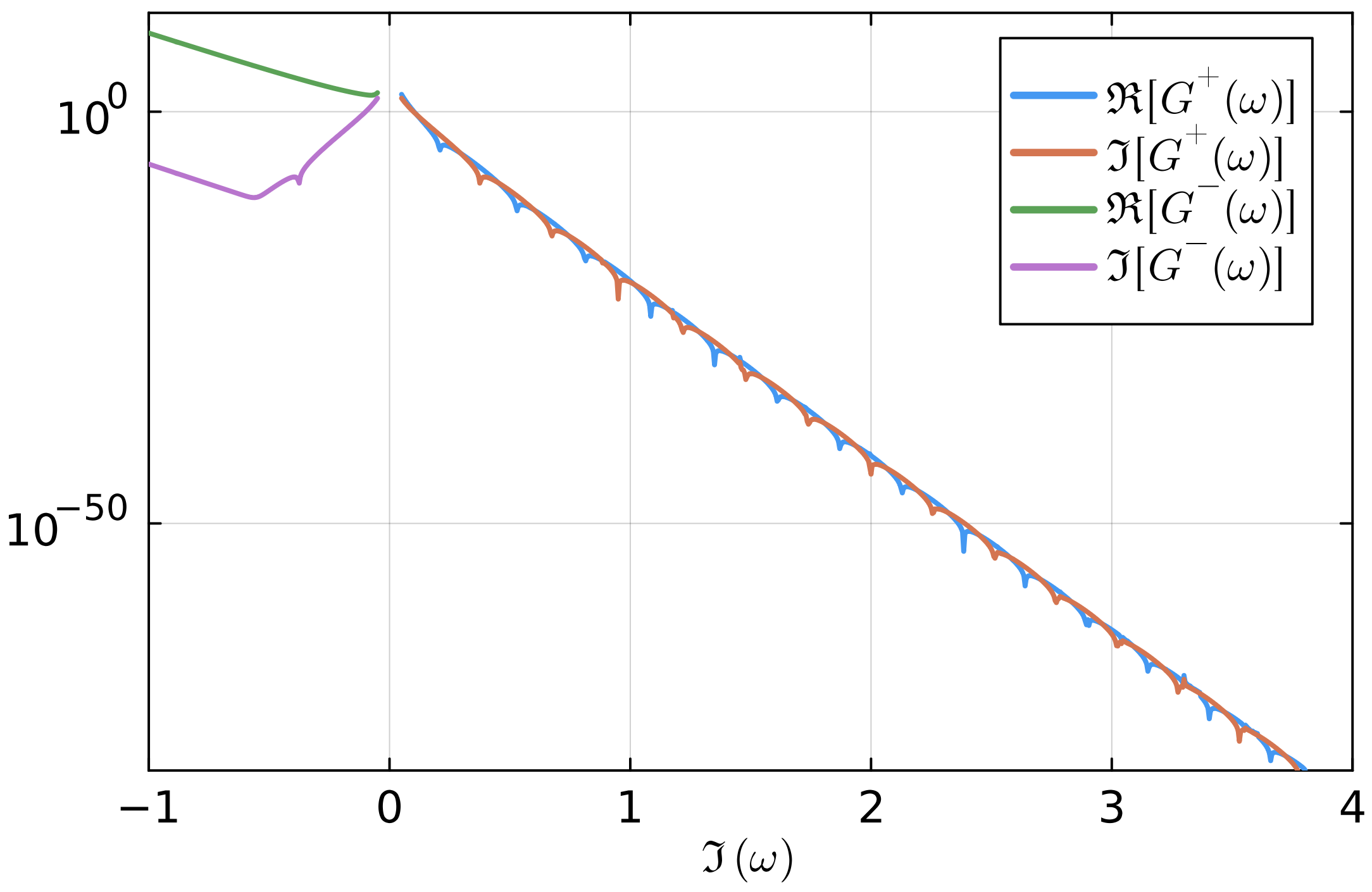}
  \caption{The frequency-domain Green’s function evaluated along the right-hand side of the BC on the imaginary-frequency axis. For positive (negative) imaginary frequencies, the curves show the real and imaginary parts of $G^{+}$ ($G^{-}$).
  }
  \label{fig:Freq_IA}
\end{figure}

\begin{figure}[t]
  \centering
  \includegraphics[width=\linewidth]{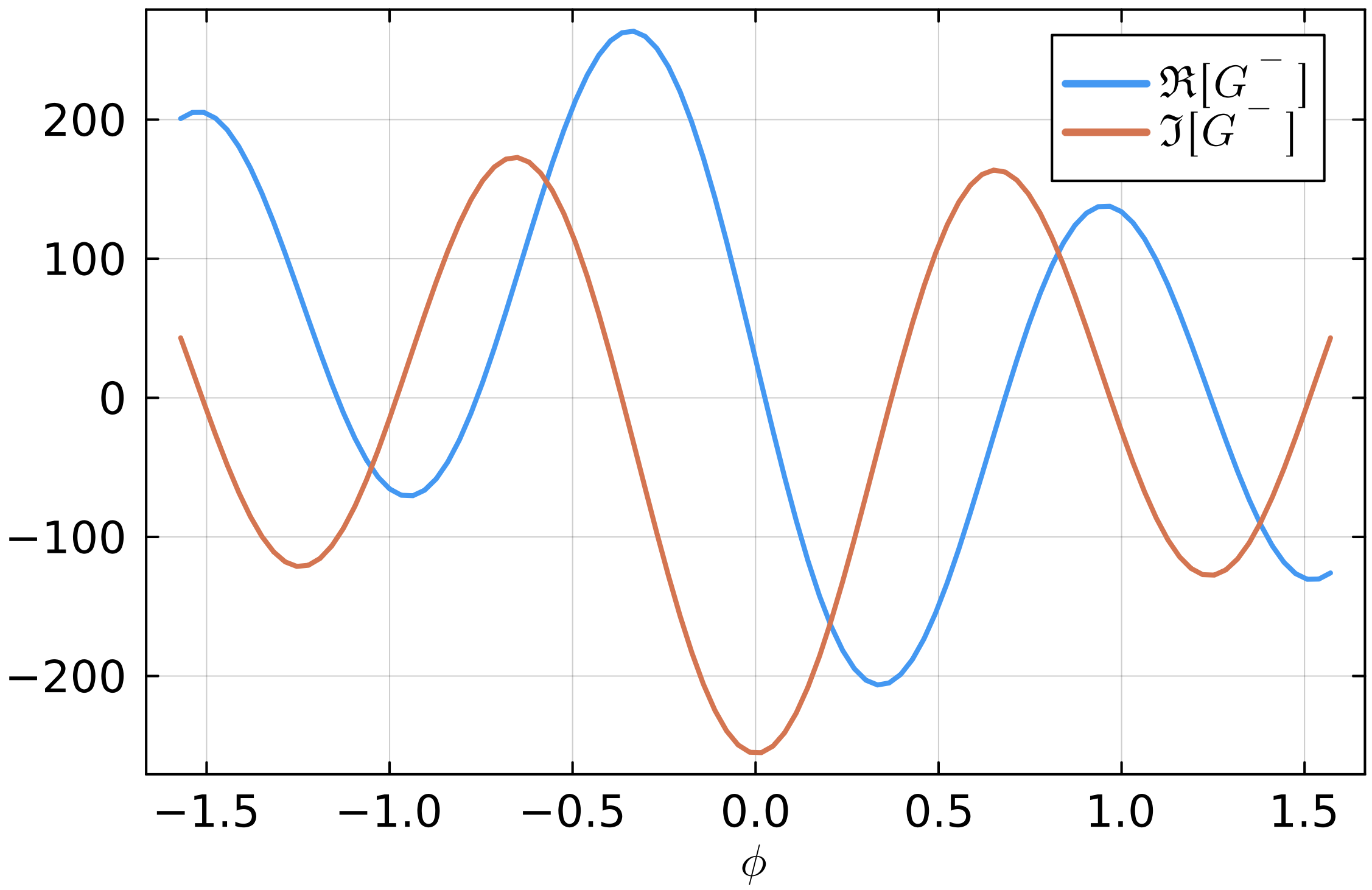}
  \caption{$G^{-}$ evaluated along the right-hand side of the small circular contour. The radius of the circle is $\delta=0.05$. The definition of the angular parameter $\phi$ is the azimuthal angle of the circular contour, as illustrated in Fig.~\ref{fig:Contour}. Here $r=30$, $r^\prime=10$, and $\ell=2$.}
  \label{fig:Freq_Small_Circle}
\end{figure}

Figure~\ref{fig:Freq_IA} shows $G^{+}$ evaluated along the right-hand side of the BC on the PIA, together with $G^{-}$ evaluated along the NIA, for $r=30$ and $r^\prime=10$. Owing to the symmetry in [Eq.~\eqref{ConjSym}], a nonvanishing imaginary part signals a discontinuity of $G^{+}$ ($G^{-}$) across the PIA (NIA).
Figure~\ref{fig:Freq_Small_Circle} shows $G^{-}$ evaluated along the right-hand side of the small circular contour illustrated in Fig.~\ref{fig:Contour}.

The direct part contribution to the Green’s function is obtained from the contour integral 
\begin{align}
G_{\mathrm{dir}}(t;r,r^\prime) 
&=
\frac{1}{\pi}
\int_{\delta}^{\infty}
\Im\!\left[ \tilde{G}^{+}(i x + \epsilon; r, r^\prime) \right]
e^{x t}\, dx
\nonumber \\
&
+
\frac{1}{\pi}
\int^{-\infty}_{-\delta}
\Im\!\left[ \tilde{G}^{-}(i x + \epsilon; r, r^\prime) \right]
e^{x t}\, dx
\nonumber \\
&+\frac{i\delta}{2\pi} \int^{\pi/2}_{-\pi/2} 
  \tilde{G}^{-}(\delta e^{i\phi}; r, r^\prime)
  e^{-i\delta e^{i\phi}t} e^{i\phi}
  d\phi
\nonumber \\
&+\frac{i\delta}{2\pi} \int^{3\pi/2}_{\pi/2} 
  \left[
  \tilde{G}^{-}(\delta e^{i(\pi-\phi)}; r, r^\prime)
  \right]^*
  e^{-i\delta e^{i\phi}t} e^{i\phi}
  d\phi
\label{DirectPartIntegration}
\end{align}
where $\Im$ denotes the imaginary part.

In the numerical implementation, radial-coordinate-independent quantities—such as the renormalized angular momentum and the asymptotic amplitudes—are precomputed and stored in a cache. For given values of $r$ and $r^\prime$, the radial-dependent parts are then evaluated efficiently. The frequency-domain Green’s function is interpolated, and the resulting interpolation function is used as the integrand in [Eq.~\eqref{DirectPartIntegration}]. The same strategy is applied in the computation of the tail and QNM waveforms. The results of our calculations are presented and discussed in Sec.~\ref{SEC:Result}.

\subsection{Tail and QNM} \label{tailQNM}
As discussed above, the tail originates from the BC of the Green’s function $\tilde{G}(\omega; r,r^\prime)$ along the NIA. The corresponding BC contribution to the time-domain Green’s function can be written as~\cite{PhysRevD.86.024021,leung2003linearizedperturbationsblackhole}
\begin{equation}
  G_B (t;r,r^\prime)
  =
  -\frac{i}{2\pi}
  \int_0^{\infty} d\sigma\,
  \tilde{G}_B (\sigma;r,r^\prime)\,
  e^{-\sigma t},
\end{equation}
where
\begin{equation}
  \tilde{G}_B (\sigma;r,r^\prime) = -2 i \sigma\,
  R^{\mathrm{in}}_{-i\sigma}(r)\,
  R^{\mathrm{in}}_{-i\sigma}(r^\prime)\,
  \frac{q(\sigma)}{|W|^2},
  \quad
  (\sigma \equiv i\omega).
\end{equation}
Here, $q(\sigma)$ denotes the branch-cut strength, which is defined through~\cite{PhysRevD.86.024021,leung2003linearizedperturbationsblackhole}:
\begin{equation}
    \Delta R^{\mathrm{up}}_{-i\sigma}(r) = i q(\sigma) R^{\mathrm{up}}_{i\sigma}(r)
\end{equation}
The quantity $\Delta R^{\mathrm{up}}_{-i\sigma}(r)$ represents the discontinuity of the UP solution across the BC, i.e., the difference between its values on the right- and left-hand sides of the NIA. Similar to the Wronskian, the branch-cut strength $q(\sigma)$ is independent of the radial coordinate $r$, and can therefore be evaluated at any convenient radius, without loss of generality.

In this work, the radial IN solution $R^{\mathrm{in}}_{\omega}(r)$ is computed using the Jaff\'{e} series representation~\cite{Leaver1986,PhysRevD.87.064010},
\begin{equation}
R^\text{in}_\omega (r) = 
\left(
       \bar{r}-1
\right)^{- i \bar{\omega}}
\bar{r}^{2i\bar{\omega}}
e^{i\omega r}
\sum^\infty_{n=0} b_n \left( 1-\frac{1}{\bar{r}} \right)^n 
    \label{Jaffe}
\end{equation}
where $\bar{r} = r/(2M)$ and $\bar{\omega} = 2M\omega$.

In the high-frequency regime, the branch-cut strength $q(\sigma)$ and the Wronskian $W$ are evaluated using the Leaver $U$-series~\cite{Leaver1986,PhysRevD.87.064010}.
The corresponding UP solution of the Regge–Wheeler equation is given by
\begin{align}
    &R^\text{up}_\omega (r)
    =
    \bar{r}^{1+s} 
    \left(\bar{r} - 1\right)^{-i\bar{\omega}} e^{i\omega r} \times
    \nonumber \\
    & 
    \sum_{n=0}^{\infty} 
     b_n \cdot (-2i\bar{\omega} + 1)_n \ U(s + 1 - 2i\bar{\omega} + n, 2s + 1, -2i\omega r),
     \label{LeaverU}
\end{align}
where $(a)_n$ denotes the Pochhammer symbol.
The expansion coefficients $b_n$ appearing in both the Jaff\'{e} and Leaver $U$-series satisfy the three-term recurrence relation
\begin{align}
\alpha_n b_{n+1} + \beta_n b_n + \gamma_n b_{n-1} = 0, \quad n = 1, 2, \ldots,
\end{align}
with
\begin{align}
\alpha_n &\equiv (n + 1)(n - 2\bar{\sigma} + 1) \nonumber\\
\beta_n &\equiv -[2n^2 + (2 - 8\bar{\sigma})n + 8\bar{\sigma}^2 - 4\bar{\sigma} + \ell(\ell + 1) + 1 - s^2] \nonumber \\
\gamma_n &\equiv n^2 - 4 \bar{\sigma} n + 4\bar{\sigma}^2 - s^2.
\end{align}
where $\bar{\sigma} = 2M\sigma$.
For the Jaff\'{e} series, the first term of the expansion coefficient is normalized to 
$b_0 = e^{-2i\bar{\omega}}$, while for the Leaver-U series, 
$b_0 = (-2i\bar{\omega})^{s+1-2i\bar{\omega}}$ ~\cite{PhysRevD.87.064010}.

In the low-frequency regime, the branch-cut strength and the Wronskian are instead computed using the MST method, since the Leaver $U$-series exhibits very slow convergence along the NIA at small frequencies.

For the QNM contribution, the Green’s function can be written as
\begin{equation}
  G_{\mathrm{QNM}}(t;r,r^\prime)
  =
  \sum_{n}
  \frac{
    A^{\mathrm{ref}}_{\mathrm{in}}(\omega_n)
  }{
    2 \omega_n\,
    \left.
    \dfrac{d A^{\mathrm{inc}}_{\mathrm{in}}(\omega)}{d\omega}
    \right|_{\omega=\omega_n}
  }
  R^{\mathrm{in}}_{\omega_n}(r)\,
  R^{\mathrm{in}}_{\omega_n}(r^\prime)\,
  e^{-i \omega_n t},
\end{equation}
where the sum runs over the quasinormal-mode overtones labeled by $n$.
The QNM frequencies $\omega_n$ are obtained using the \texttt{qnm} module of the \texttt{BlackHolePerturbationToolkit}~\cite{Stein:2019mop}.
The asymptotic amplitudes are computed using our MST implementation, while the ingoing radial solutions evaluated at the QNM frequencies are obtained from the Jaff\'{e} series representation.
The derivative of the incident amplitude $A^{\mathrm{inc}}_{\mathrm{in}}(\omega)$ is computed numerically using Richardson extrapolation~\cite{Lo:2025njp}.

\subsection{Time-domain simulation} \label{Simulation}

To validate the reliability of our frequency-domain calculations, we perform an independent time-domain simulation for comparison. The simulation is carried out on a uniform grid in the tortoise coordinate $r_*$. For observers located at finite radius, the computational domain in $r_*$ is chosen to be sufficiently large so that spurious boundary effects do not contaminate the waveform within the time interval of interest. Time evolution is performed using a second-order leapfrog finite-difference scheme.

To approximate the delta-function source in both the radial and temporal directions, we employ a narrow Gaussian profile,
\begin{equation}
  \delta(\mathbf{x})
  \;\approx\;
  \frac{1}{2\pi \sigma_{r_*}\sigma_t}
  \exp\!\left[-\frac{(r_* - r_*^\prime)^2}{2\sigma_{r_*}^2}\right]
  \exp\!\left[-\frac{(t - t^\prime)^2}{2\sigma_t^2}\right],
\end{equation}
and impose vanishing initial conditions for both the field $\psi$ and its time derivative $\partial_t \psi$.

In cases where the late-time tail signal is contaminated by numerical noise, we perform the simulation using 128-bit floating-point arithmetic provided by the \texttt{DoubleFloats.jl} Julia library, employing the \texttt{Double64} type throughout the computation to enhance numerical precision. We further carry out convergence tests with respect to the Gaussian widths, the time step, and the radial grid spacing. These tests confirm that the time-domain simulation provides a reliable approximation to the Green's function.

For the simulations shown in Figs.~\ref{fig:BCDP_early}, \ref{fig:Filter}, and \ref{fig:FullGF}, the grid spacing is chosen as $\Delta r_* = 0.01$, with the computational domain extending from $r_*=-1000$ to $r_*=1000$. The time step is set to $\Delta t = 0.002$, and the Gaussian widths are taken to be $\sigma_{r_*}=0.1$ and $\sigma_t=0.1$.

\section{Results} \label{SEC:Result}

\begin{figure}[t]
  \centering
  \includegraphics[width=\linewidth]{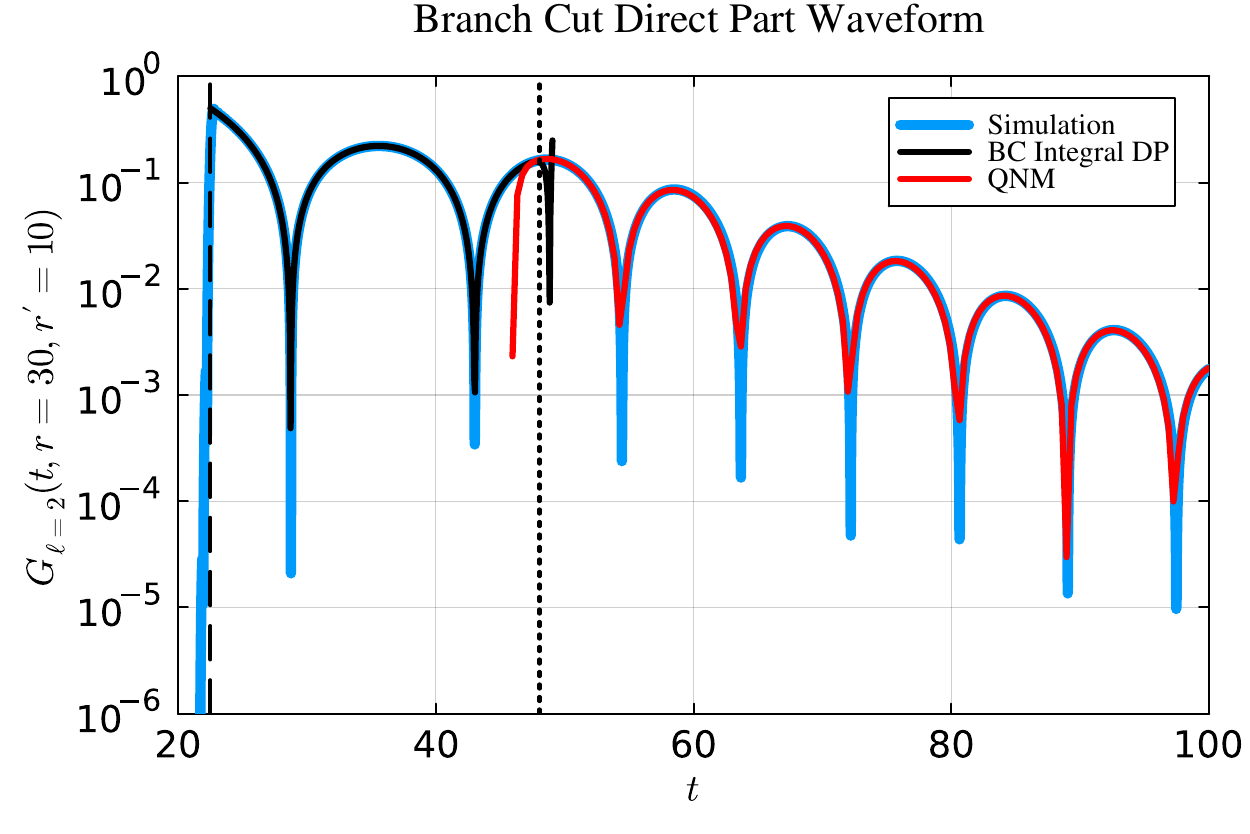}
  \caption{Early-time behavior of the time-domain Green's function.
The blue dashed curve shows the full Green's function obtained from the time-domain simulation.
The orange curve represents the BC direct part,
while the green curve corresponds to the sum of QNMs with overtones $n=0$--$7$.
The black dashed line indicates the time $t = r_* - r^\prime_*$, and the black dotted line indicates $t = r_* + r^\prime_*$. 
This convention is adopted for all subsequent figures. In this figure, $r=30$, $r^\prime = 10$ and $\ell = 2$.}
  \label{fig:BCDP_early}
\end{figure}

In this section, we present the results of our calculations.
Fig.~\ref{fig:BCDP_early} compares the direct part waveform and the QNM waveform with the time-domain simulation.
The direct part waveform shows excellent agreement with the simulated waveform in the time interval
$t \in ( r_* - r_*^\prime , r_* + r_*^\prime )$.
At later times, $t > r_* + r_*^\prime$, the time factor $e^{-\sigma t}$ in the PIA BC integrand is no longer sufficient to suppress the contribution from the frequency-domain Green’s function $G^+$ along the PIA, and the corresponding integral becomes nonconvergent.
In this regime, the BC contour used to isolate the direct part is no longer valid.
Instead, the traditional Leaver contour which naturally incorporates both the QNM contributions and the BC tail provides the appropriate description of the waveform.

\begin{figure}[t]
  \centering
  \includegraphics[width=\linewidth]{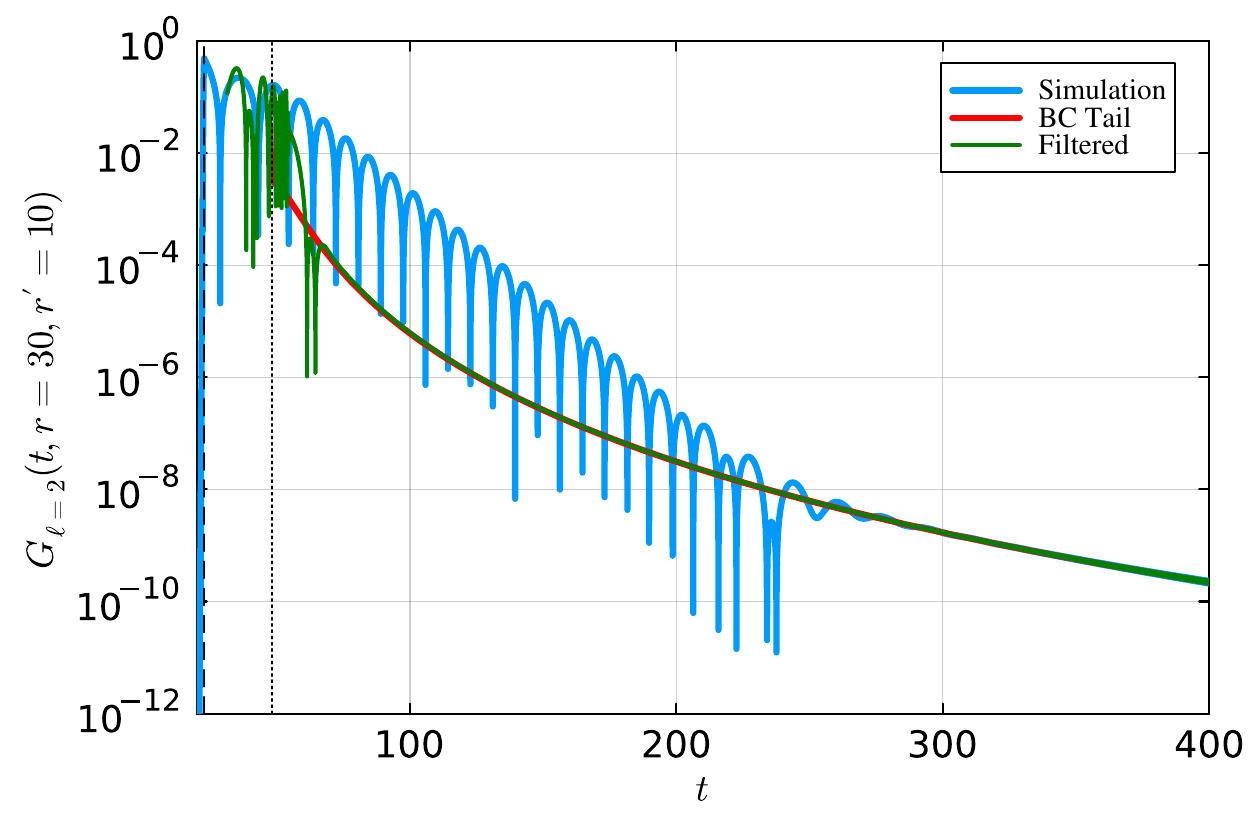}
  \caption{
  Comparison between the BC tail contribution and the numerically simulated waveform after applying the QNM filter described in~\cite{PhysRevD.106.084036}. In this figure, $r=30$, $r^\prime = 10$ and $\ell = 2$.
  }
  \label{fig:Filter}
\end{figure}

The red curve in Fig.~\ref{fig:Filter} represents the tail waveform obtained from the NIA BC integral.
At late times, it agrees well with the tail behavior observed in the time-domain simulation.
However, during the QNM-dominated phase, the tail contribution is masked by the QNM oscillations, making a direct comparison difficult.
To disentangle the tail signal, we apply the recently developed QNM filter proposed by Ma \textit{et al.} to remove the $n=0-7$ overtones from the simulated waveform (see Refs.~\cite{PhysRevD.106.084036,PhysRevD.107.084010,PhysRevLett.130.141401} for details).  See~\cite{Baibhav:2023clw} and references therein for alternative approaches.
The resulting filtered waveform, shown as the green curve in Fig.~\ref{fig:BCDP_early}, exhibits excellent agreement with our theoretical prediction.
This agreement provides a strong validation of our calculation.

\begin{figure}[t]
  \centering
  \includegraphics[width=\linewidth]{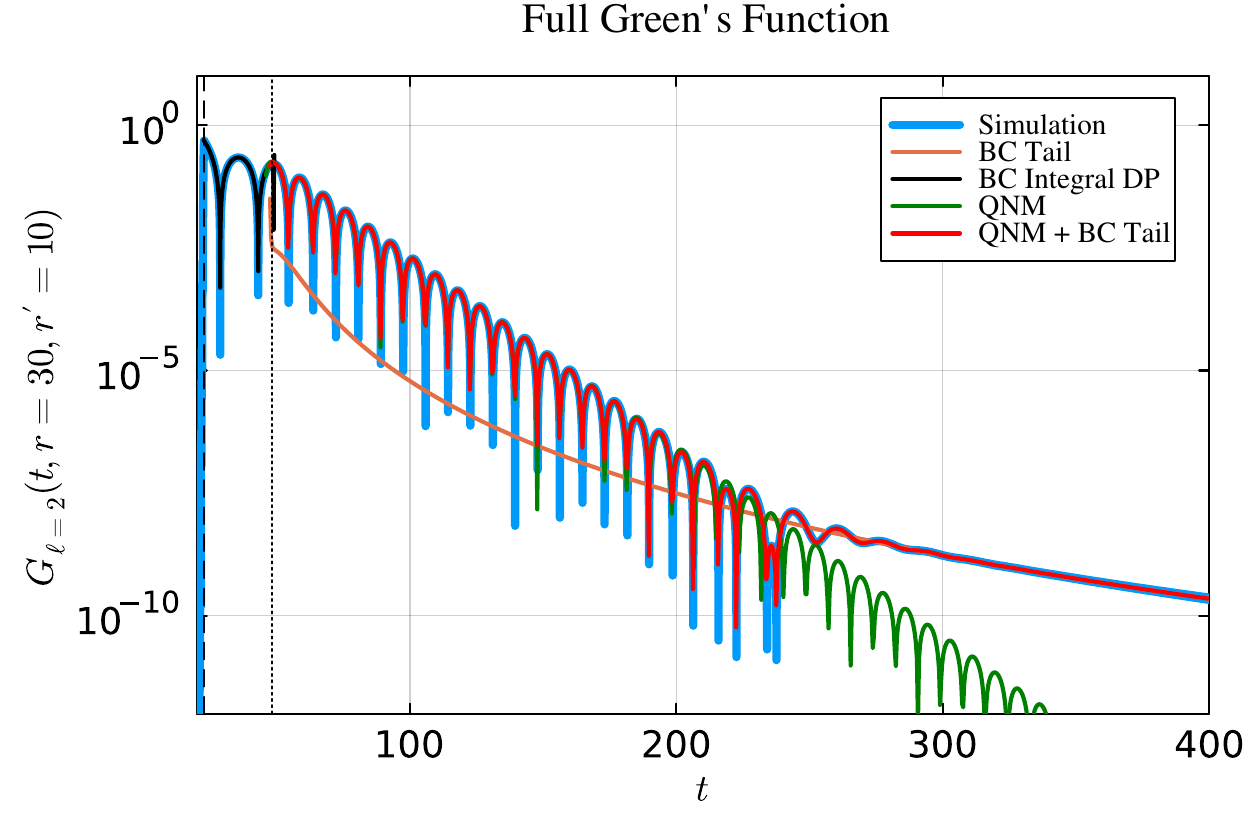}
  \caption{Comparison between the full contour-integral result and the time-domain numerical simulation of the Green’s function.
The blue curve shows the simulated waveform.
The black, green, and orange curves correspond to the direct part, the QNM contribution, and the BC tail, respectively.
The red curve represents the sum of the QNM and tail contributions.
In this figure, $r=30$, $r^\prime = 10$ and $\ell = 2$.
}
  \label{fig:FullGF}
\end{figure}

\begin{figure}[t]
  \centering
  \includegraphics[width=\linewidth]{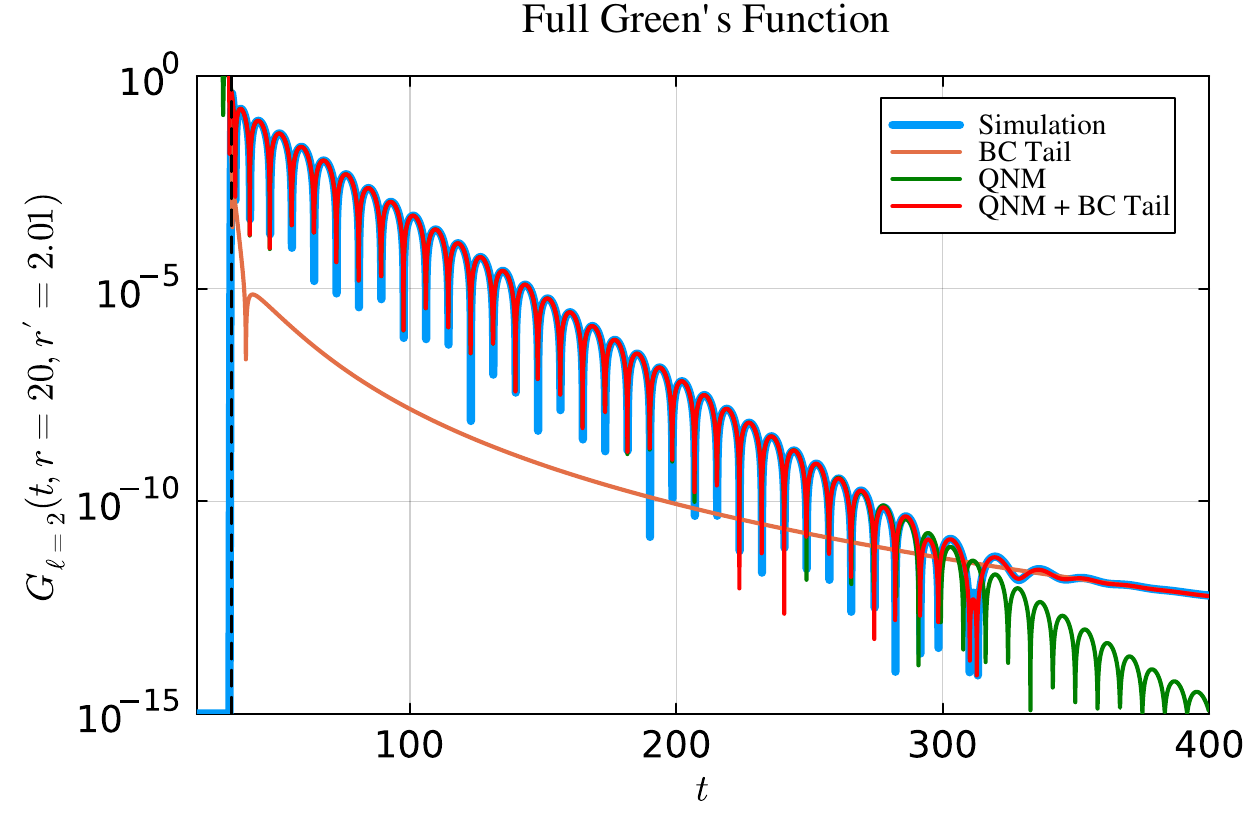}
  \caption{
Comparison between the full contour-integral result and the time-domain numerical simulation of the Green’s function. The blue curve shows the simulated waveform, while the green and orange curves represent the QNM contribution and the BC tail, respectively; the red curve denotes their sum. Here \( r = 20 \), \( r' = 2.01 \), and \( \ell = 2 \). Since \( r_*' < 0 \), no direct part contribution is present.
}
  \label{fig:FullGF2}
\end{figure}

Notice that the tail waveform diverges for $t < r_* + r_*^\prime$, similar to the direct part regime.
The frequency-domain Green's function (both its QNM and NIA BC contributions) also exhibits a behavior
$\sim e^{\sigma (|r_*| + |r_*^\prime|)}$, aside from subdominant factors behaving polynomially in $\sigma$,
as shown in Ref.~\cite{PhysRevD.86.024021} (this readily follows from Eqs.~9, 37, 46 and 57 therein), which becomes $\sim e^{\sigma (r_* + r_*^\prime)}$ in the setting here where  $r_*,r_*^\prime>0$.
For $t < r_* + r_*^\prime$, the time factor $e^{-\sigma t}$ no longer suppresses the frequency-domain Green’s function to yield a convergent integral, rendering the NIA BC contour invalid in this regime.

Figure~\ref{fig:FullGF} shows the direct part, the QNM contribution, the tail contribution, and the time-domain numerical simulation plotted together. We also display the waveform obtained by summing the QNM and tail contributions, which begins at \( t = r_* + r_*^\prime \) and agrees precisely with the numerically simulated waveform.

Figure~\ref{fig:FullGF2} illustrates the case \( r_*^\prime < 0 \), where we take \( r^\prime = 2.01 \) as the radial coordinate of the source. As in Fig.~\ref{fig:FullGF}, the QNM and tail contributions are shown. In this case, however, the QNM and tail contributions alone are sufficient to reproduce the Green’s function waveform, as the direct part is absent, consistent with the analysis presented above.
These results demonstrate that the full waveform of the Schwarzschild Green’s function can be accurately reconstructed from first principles using our contour-integral formulation.


During the preparation of the current manuscript, we have noticed a recent work in \cite{deamicis2026postminkowskianexpansionpromptresponse} that analyzed Schwarzschild Green's function in the Post-Minkowski limit. Although in our analysis the direct part comes from both the small arc around $\omega=0$ and the BCs on the imaginary axis, it is reasonable that in the limit considered in \cite{deamicis2026postminkowskianexpansionpromptresponse} it is dominated by the contribution of the $\omega=0$ branch point.

We also note that the $\ell=2$ Green function of the Regge-wheeler equation was  obtained in~\cite{otoole2020characteristic} via a numerical method using characteristic initial data, as opposed to via a spectroscopical decomposition as done here. In~\cite{aruquipa2026greenfunctionsreggewheelerteukolsky}, the $\ell=2$ Green function of the Teukolsky equation was  obtained   by Fourier-integrating along the real frequencies and it was compared to the dominant QNM and NIA BC contributions (without including a direct part as done here; the full $4$-D Green functions of the Regge-Wheeler and Teukolsky equations were also numerically calculated in ~\cite{aruquipa2026greenfunctionsreggewheelerteukolsky}).


\section{Conclusion}
In this work, we have completed the theoretical framework to individually define and compute the direct part, the QNMs, and the tail of a Schwarzschild spacetime. The key innovation with respect to the exercise in the Schwarzschild-dS spacetime \cite{arnaudo2025quasinormalmodescompletemode}, is to replace the contribution of the Matsubara modes by the BCs on the imaginary axis. In this framework, the direct part is contributed by the BC integrals on both the positive and negative imaginary axes, and a small arc around the branch point at $\omega=0$. We expect that similar analytical structure should hold true for the Kerr Green's function. However, the split in the Kerr case is considerably more involved than in the Schwarzschild geometry. 

First, the Kerr Green’s function possesses fewer symmetries than Schwarzschild. As a consequence, a significantly larger number of independent quantities must be computed in order to evaluate the Green’s function along the BCs and the small circular contour in the complex frequency plane. Second, the validation of time-domain results is substantially more challenging in the Kerr case. In Schwarzschild spacetime, different $\ell$-modes are decoupled, and the time-domain evolution reduces to a one-dimensional problem, making independent numerical simulations and
comparisons relatively straightforward. In contrast, Kerr perturbations involve coupled spherical $\ell$-modes, as well as a mismatch between the angular bases of spin-weighted spherical harmonics and spheroidal harmonics, which
significantly complicates time-domain simulations and their comparison with frequency-domain calculations.


Despite these challenges, a quantitative decomposition of a Kerr Green's function will be very useful, since its application to plunging particles will give us a robust prediction of the dynamically excited QNMs, dynamically excited direct wave, and the tail. While the early part of the tail seems to be sub-dominant, both the direct wave \cite{Oshita:2025qmn,Lu:2025vol} and the QNMs are directly relevant for constructing a ringdown waveform. This linear decomposition of signals can be further used to compute nonlinear signals at second and higher orders as consequences of gravitational wave-wave couplings \cite{Mitman:2022qdl,Cheung:2022rbm,Khera:2023oyf, Cheung:2023vki,Ma:2024qcv,Bourg:2024jme,Khera:2024bjs,Wang:2026rev,Sberna:2021eui, Bucciotti:2024zyp, Lagos:2022otp}, which are also important targets to search for in the black hole ringdowns.

\begin{acknowledgments}
This work makes use of the Black Hole Perturbation Toolkit.
HY is supported by the Natural Science Foundation of China (Grant 12573048). NK is supported by the Shuimu fellowship of Tsinghua University. 
 SM acknowledges support from the Natural
Sciences and Engineering Research Council of Canada
through a Discovery Grant. This research was supported in part
by Perimeter Institute for Theoretical Physics. Research
at Perimeter Institute is supported in part by the Government of Canada through the Department of Innovation,
Science and Economic Development and by the Province
of Ontario through the Ministry of Colleges and Universities. AC is funded by Beijing Natural Science Foundation (BNSF) International Scientists Project (Grant No. IS25033) and also from a postdoctoral fellowship (Grant
No. 202504) through the Department of Astronomy, Tsinghua University.
\end{acknowledgments}  

\bibliography{References}

\end{document}